# On the ideas of the origin of eukaryotes: a critical review

*Parsifal Fidelio Islas-Morales [abc] & Luis Felipe Jiménez-García [a]*

[a] Programa de Doctorado en Ciencias Biomédicas, Facultad de Medicina, UNAM
[b] Department of Cell Biology, Faculty of Sciences, UNAM
[c] UNESCO Chair in Science Diplomacy and Heritage, UNAM

## Abstract

The origin and early evolution of eukaryotes is one of the major transitions in the evolution of life on earth (Smith and Szathamary, 1995). One of its most interesting aspects is the emergence of cellular organelles, their dynamics, their functions, and their divergence. Cell compartmentalization and architecture in prokaryotes is a less understood complex property. In eukaryotes it is related to cell size, specific genomic architecture, evolution of cell cycles, biogenesis of membranes and endosymbiotic processes. Explaining cell evolution through form and function demands an interdisciplinary approach focused on microbial diversity, phylogenetic and functional cell biology. Two centuries of views on eukaryotic origin have completed the disciplinary tools necessarily to answer these questions. We have moved from Haeckel's *SCALA NATURAE* to the un-rooted tree of life. However the major relations among cell domains are still elusive and keep the nature of eukaryotic ancestor enigmatic. Here I present a review on state of art views of eucaryogenesis; the background and perspectives of different disciplines involved in this topic

# Contents

**Ideas in Eukaryogenesis**

Historically, there have been two reductionist perspectives: 1) one that reconstruct the history of organelles based on their biology and build up a hypothetical scenario to fit into *a priori* assumptions, and from which contradictions have arisen when trying to fit individual organelle origin into a wider model. Saltationst scenarios as those proposed by endosymbiosis contrast with gradualist scenarios as those proposed by archetypal hypothesis 2) the second perspective reconstructs the history of eukaryotic diversity based on phylogenetic hypothesis as a general picture. The disadvantages of phylogenetic algorithms are common to this kind of approach. Here we will try to contrast both perspectives into an integral view of the state of art of eukaryogensis and its disciplinary perspectives.

**Palaeobiological perspectives**

A usual tool in phylogenetic reconstruction is the use of fossil record to calibrate relaxed molecular clocks and topologies (Roger & Hug, 2006; Katz, 2012). This approach can led to a better picture of diversification times and coherence in diversification patterns. Among the resolved division between unikonts and bikonts, major lineages such as Excavata, Archeaplastida, Opistokonta, Amebozoa and SAR clade reveal strong basal divergence (Adl *et al.,* 2012). However phylogenetic relations and divergence sequence have not been resolved. The question about date and historic narrative of the crown group of eukaryotes can contrast with the many hypotheses about origin of eukarytoes (Knoll, 2014).

The fossil record can support radiation and diversification patterns, for example known eukaryotic diversification of certain lineages during the Phanerozoic. Crowngroup diversification should be traced back to the Neoproterozoic. The Proterozoic (542-2500 Mya) and the Archean (2500-4567 Mya) might have evidence on the nature of eukaryotic cells and the kind of environment; eukaryotic first diversification took place (Knoll, 2014). Cell like fossils have been found as vesicles in 3500 Mya old rocks in the Archean (Javeaux, 2011; Butterfield, 2004). Bangiomorpha, a Rodophyte like, has been dated between 1100-1200Mya (Butterfied 2000). Laminar wall structure attributed to chlorophytes have been dated in mid-Proterozoic strata (Javeaux, 20004). Shells are also abundant in 1600 Mya sediments, however lack of ornamental diagnostic structures. Silica and calk tests of putative amoeba and simple foraminifera are found in the mid Neoproterozoic (Porter & Knoll, 2000). Nodules and vessels comparable to extant chrisophytes are dated 800Mya (Allison & Hilgert, 1986), before Ediacarian radiation. Most putative eukaryotes are difficult to characterize due to the lack of diagnostic characters. Javaux has established general identification criteria: size, complex ultrastructure and thickness of walls and shells (Jauveax, 2003).

Complementary to marginal morphologies, molecular fossils are a tool for paleobiologists. Steranes, derivate forms of sterols, are hallmarks of possible eukaryotic sterol biosynthesis (Pawlowska *et al.,* 2013). Methanogenic activities can be traced to 2700 Mya, suggesting that eocytes hypothesis can be drawn no later than 2700 Mya (Hayes, 1994; Knoll 2014).

Gradualistic and endosymbiotic scenarios are challenged since sterol synthesis in the Archean could just take place at nanomolar levels (Holland, 2006). Steranes register from the Archean are rare and can be due to contamination (Dutkiewicz *et al.,* 2006). Chronic increase of oxygen in ocean water is fond until de Proterozoic and supported by the divergence time of mitochondrial genes for aerobic respiration. Plastid acquisition can be traced before 1300 Mya. Chronic oxygenation can be due to cyanobacteria photosynthesis on microbial mats (Ambar *et al.,* 2007).

A comparison between early eukaryotic diversification and Ediacarian fauna and Cambiran carnivory has been proposed (Gingras *et al.,* 2011, Erwin *et al.,* 2011). Evolution of phagotrophy could had triggered the process of early diversification (Porter, 2011). Phagotrophy other than predation of little bacteria is found dispersive in several eukaryotic lineages suggesting that ancestral lineages evolved due to phagotrophic selective pressures. Fossils of ciliates, dinoflagelates and amoebas with protective armor support this scenario where size gain or aggregation could have been an adaptation against phagotrophy (Cohen & Knoll, 2012). Some authors even place eukaryogensis at the in the origin of multicellularity (Erwin *et al.,* 2011).

**Theoretical perspectives**

*Endosymbiotic theory*

In 1979 Lynn Margulis, based on the ideas of Mereschowski, proposed a theory on the endosymbiotic origin of mitochondria and undulipodia. A mutualistic relationship between two bacteria would lead to the total incorporation of both functions in a single cell. Coevolution and symbiosis became the basis to say that the origin of eukaryotes was chimeric and is in essence polyphyletic (Margulis & Bermudez,1984; Sapp, 2004)..

In 1990 Carl Woose proposed the so-called "standard model" where archaea are related to eukaryotes (Woese *et al*., 1990). The one gene (16S SSU) phylogeny by Woose provided a cue for the nature of the putative symbiotic partner of ancestral mitochondria: the archaea (Pace, 2006).

In same decade, the mitochondria´s origin was found within the lineage of the α-proteobacteria, as genomic alignments of the small ribosomal subunit revealed (Sapp, 2004). During coevolution, most metabolic and structural genes have disappeared from mitochondrial genome, so alotopic expression of their sequences in the nucleus became a condition that enabled form and function of these organelles (Gonzalez-Halpen *et al*., 2003. Mitochondrial ancestry was found to differ in various eukaryotic lineages suggesting that endosymbiosis could occur often in the course of evolution (Degli Esposti, 2014). Eukaryotic features were a byproduct of symbiotic co-evolution, however early endosymbiotic theory could not show evidence for this (Lang *et al*., 1999).

*The Archeozoa hypothesis*

In contrast to endosymbiotic theory, Thomas Cavalier-Smith proposed a gradualist view of the origin of eukaryotes in 1987. He emphasized on the structural and physiological constrains that would have been necessary for endosymbiosis: such as phagocytosis, cytoskeleton. He noted also on the lack of a logical, evidence based explanation for an endosymbiotic origin of the cell nucleus. For Cavalier-Smith gradual evolution of ancestral forms is a need before endosymbiosis, which he does not, held as the main driver of eukaryogenesis.

The archeozoa hypothesis considers the engulfment of α-protobacteria by a protoeukaryotic ancestor, already capable of phagocytosis (Cavalier-Smith, 2009, Poole and Penny 1991). Eukaryotic features evolved not as consequence of mitochondrial pressures of selection, but as idiosyncratic characters. Eukaryotes evolved from a common ancestor within bacteria. The Neomuran ancestor proposed by Smith is related to extant endobacteria and was the ancestor of wall-less actinobacteria. This organism developed phagotrophy, endomembranes, mitosis, sex, nucleus, and cilium early before mitochondria were integrated by endosymbiosis (Cavalier Smith, 2001).

Cavalier Smith idea relies on cell biology, micropaleontology, and comparative protozoology. Extant organisms as *Pelomyxa palustris* and further amitochondriate Excavate were thought to be possible archeozoa living fossils (Cavalier Smith, 1991). *P. palustris* does indeed lack of mitochondria but replace them with enosymbiotic bacteria, furthermore endoplasmic reticula are absent. *G. lambia* also lack mitochondria and together with *Entamoeba hystolitica* were to be thought as basal eukaryotes. In the later years, hidrogenosomes, mitosomes were found in amitochondriates. This organelles harbor genetic hallmarks of secondary mitochondrial loss as demonstrated in *Gardia lambia* and *E. histolytica* (Dyall & Johnson 2000; Clark & Roger, 1995). Archeozoa hypothesis was partially dismissed as the idea of ancestral eukaryotes among extant representatives of amoeba and excavates could not be proved.

Endosymbiotic and archeozoa theory differed on the chronology of endosymbiotic events and thus the importance of endosymbiosis as a driver of organelle origin. They contrasted a co-evolutionary perspective against an idiosyncratic scenario.

*The chimeric origin of the nucleus*

The late work of Margulies proposed that the nucleus arose before the mitochondrial endosymbiosis as the result of symbiosis between an archaeal thermoplamsa (eocyte) and a eubacterial spirochete (Margullis *et al.,* 2010). Eocyte, spirochete and sulfur globules developed a syntrophic strategy as adaptive trait. The microbial community, *Thiodendron latens,* offers an extant example of such a consortium where sulfate reduction is provided as source of electron acceptors for spirochetes that provide for efficient carbohydrate metabolism or for endosymbiotic motility related benefits (Dubinina *et al.,* 2004). DNA exchange must have taken place. Vesicle formation due to membrane hypertrophies can take place as seen in *Gemmata obscuriglobis* (Lindsay *et al.,* 2001). This combination gave rise to the cytological structure known as karimastigont as direct ancestral state of the later eukaryotic nucleus. This structure converted into primitive microtubuli organizing center. Afterwards nuclei separated from MTOC and toke a central position in the cell, as seen in many early branching protest (parabasalids and hipermastiginds) (Margullis *et al.,* 2010). The extant ciliate *Mixotrichia paradoxa* harbor thousand of anchored spirochetes for motility purposes (Wenzel *et al.,* 2003).

*Viral Eukaryogensis*

The lack of transitional organisms between prokaryotes and eukaryotes has driven to speculative approaches additionally to the efforts of the genomic era. The viral eukaryogensis hypothesis was proposed by John Bell in 2001 based on a consortium theory where archaea, bacteria and complex DNA virus form a gradual symbiosis which results in the chimeric nature of eukaryotic cell (Bell, 2001). The main concerns of viral eukaryogensis are the gap between complexity in the eukaryotic cell cycle and genome architecture; and the prokaryotic features (Bell, 2009).

The hypothesis is based on a theoretical experiment on biological actualism. If the origin of eukaryotes took place in a prokaryotic world, then the features or a chimeric ancestor of eukaryotes can be found in extant representatives of prokaryotic domains, including viri. Membrane and cytoplasm as known in eukaryotes are proposed to origin from archaeal wall-less methanogens as *M. eliziabethi* (Rose & Pirt, 1981). The syntrophic nature of methanogen in microbial consortia opens the possibility for them to establish symbiotic relation with other bacteria (Bell, 2009). This idea is consistent also with the argument in favor for an Archaean origin of the cytoskeleton and the membrane remodeling complex, despite these implications have not been discussed extended in the context of viral eukaryogensis (Koonin, *et al.,* 2006).

The most innovative element of viral eukaryogensis is the suggestion of the nucleus originated from a complex DNA virus. There is certainly molecular, cytological, and biochemical basis that argue for the genome a physiological complexity of DNA viri; especially giant viri form the NCLDV group (Villareal *et al.,* 2000; Bell, 2009). Mimiviri, pandoraviri and poxviri have been found capable of double strand DNA replication inside their capsid. Thus, their replication mechanism, in hyperpolyploid hosts, is the generation a viral factory. RNA capping is also a feature of NCLDV viri, suggesting that posttranscriptional modifications could have started in viral factories (Raoult *et al*., 2004; Koonin & Yutin, 2010).

In *Acantoamoeba polphaga:* mimiviri, rickettsias, nuclei and mitochondria coexist. Taking this as inspiration, a consortium of wall-less archaea, α-protobacteria and large DNA viri is plausible (Horn & Wagner, 2004). Giant viri normally set up to lysogenic phase as this ensure the protection their DNA within the host and coordinated propagation. Bacteriophage P1 like virus and recently described archaeal viri (Forterre. 2012) can be considered triggers of membrane evolution (Bell, 2013). Bacteriophages develop vesicles from bacterial membrane invagination, and viral factories need of vesicle formation to build new membrane bound capsids (Karhu *et al,* 2007). In this hypothesis viral factories also may have overtaken the protagonist role of protecting DNA. Giant viri acquire genome complexity through alotopic transfer of host genes. Fine gene regulation inside virus membrane could have offered an advantage in comparison to free swimming chromatin (Bell, 2013).

The conclusive remark of viral eukaryogensis is that it proposes a model of evolution of mitosis and sex in early eukaryotes. Mitosis can arise from mechanisms for chromosome *btw.* plasmid´s segregation. Conjugation could have arisen as a process triggered by predation via phagocytosis. Viral nuclei with certain genetic signatures can exchange information. Finally, meiosis and sex can be seen as consequences of errors in control mechanism for ploidy, and advantage for improving genetic diversity within the populations (Bell, 2006).

*The communities and ecosystem approach*

The ecosystem approach has been poorly integrated to the state of art of eukaryotic origins because of our ignorance on the nature of Proterozoic micro-ecologies. Paleobiology and astrobiology as scientific programs have given cues to ecosystem physical conditions and main ecological triggers in early evolution.

Recently some theoretical approaches to complex systems and communities' behavior have been argued towards an ecosystem first theory of the origin of life and eukaryogensis. Norris and Root-Bernstein propose that the eukaryotic cell originate from the integration of hyperstrucutres in prokaryotic cells based on the principle of molecular complementarity (Norris & Root-Bernstein, 2009). For them, eukaryotic features are the result of a process of progressive lost through selection of efficient ecologies, rather than individual traits. Systems theory establish that the elements of a complex system evolve in the context of a common ecology (Hunding *et al.,* 2006). Furthermore, ecologies give rise to new ecologies. Departing from the ontological question that the unrooted tree of life sets: ¿What was the nature of LUCA? The authors argue on a scenario were the evolution of molecular ecologies give rise to highly complex protocells *btw.* protecosystems. Later, a progressive loss of complexity gives rise to individual cells (Norris *et al.,* 2007). The origin of eukaryotes can be traced to their emancipation from complex molecular ecosystems (Norris & Root-Bernstein, 2009).

**The Bioenergetics based scenarios**

Metabolism based hypothesis suggest that Eukaryotes arose from the invasion of an archaeon by a protobacteria (Martin & Müller. 1998; Rivera & Lake, 2004; (López-García & Moreira, 1998). This could have been the start point for metabolic selection pressure that conformed either metabolic compartmentalization or DNA isolation through the nucleus. Following hypothesis argue in favor of this with consequences on the context of accepted phylogenetic inferences.

*Quantitative approaches*

Bioenergetics hypotheses argue that endosymbiotic events provided new orders of magnitude of energetic demand (Lane & Martin, 2010). In words of Nick Lane an important energetics divide is fond between prokaryote and eukaryotes. The last expanded their genome, expression and metabolic surface over 200,000-fold. The hypothesis is that endosymbiosis with facultative anaerobic α-proteobacteria resulted in genomic asymmetry by accumulation of bacterial genomes and metabolic activity. The demand of increased space and gene expression was solved by mitochondrial coordination that ensures form expanded metabolic surface (Lane, 2011). Examples for expanded metabolic surface can be seen in bacterial gigantism. Those organisms are hyperpolyploid and solve genetic asymmetry through metabolic membrane sustainability (Soppa, 2014).

*The hydrogen hypothesis* (Martin & Müller. 1998)

In the hydrogen hypothesis, eukaryotes are a result of spontaneous symbiosis between a euryarchaeal methanogen and facultative anaerobic eubacteria. It is also a syntrophic relation were first anaerobic respiration in eubacteria produces hydrogen, acetate and $CO_2$ as products which the archaeon uses for methanogenesis. In potentially anoxigenic media such as the reducing atmosphere of the Archean, dependence between the two types could lead to engulfment of the bacteria by the wall-less euryarchaeota. Endosymbionts could have kept genes for aerobic respiration, leading to different kinds of eukaryotes depending on adaptive radiation triggered by oxygen availability. Amitochondriate with mitoses result in fermentations, amitochondriate with hydrogenosomes result from reversion to PDH metabolism (Martin & Müller. 1998). Main disadvantage of hydrogen hypothesis is that current phylogenies tend to dismiss the idea of euryarchaeal origin.

*The syntrophic hypothesis*

The syntrophic hypothesis proposes that euryarchaeal methanogens coexist with δ-proteobacteria (ancient myxobacteria) with whom they enter a primary endosymbiosis in the Archean eon (Moreira & López-García, 1998). Fermentative sulfate reducing δ-proteobacteria provided the methanogen the selection pressure for building up sulfate reduction metabolic compartments. As the result of this endomembrane, probably nucleus arose. α-proteobacteria endosymbiosis was secondary and led to evolution of mitochondria after oxygenation of oceans in the Proterozoic (López-García & Moreira, 2006).

*The eocyte hypothesis*

The eocyte hypothesis was proposed by Lake as phylogenetic explanation of eukaryogenesis in accordance with the standard model of Woose. If ribosomal sequences place an un-rooted tree domain tree of life, Eukaryotes should have evolved from extant Crenarchetoa (Eocytes) as they share most conserved genes (Rivera and Lake 2004). The nature of this evolution was elusive. Eukaryotes could be rooted within the Crenarchaea or in a discontinuous scenario, endosymbiotic events happened to members of Crenarchaea (Eocyte) which gradually converted into modern eukaryotes. Phylogenetic technical advances such as models that reduce long branching attractions and fast evolving artifacts were considered to this hypothesis (Cox *et al.,* 2008). However, the phylogenies resulted in elusive polytomies. Archaea groups were not enough to provide strong signal and solve the phylogeny, which remained enigmatic.

**From Carl Woose´s standard model to the comparative genomic diversity of Archaea**

Improvement of phylogenetic studies took place in the following years by concatenation of highly conserved proteins, comparison of phyletic patterns, and comparison of protein domain architectures. They showed that eukaryotes harbor bacterial and archaeal homologues in addition to idiosyncratic genes. The identification of these genes and domains helped to understand their story as products of lateral gene transfer or vertical inheritance and relate it to their function. Bacterial homologues in eukaryotes correspond to functional operational characters and bioenergetics processes, while archaeal homologs correspond to fundamental information flow and housekeeping tasks (Koonin, 2015; Cox *et al.,* 2008). Despite efforts in phylogenetic discrete topologies between Eukarya and Archaea were elusive until the 2010. Vellai argued that, according to the standard model, the ancestral eukaryotes are placed outside the extant archaeal diversity (Vellai *et al*., 1998). This panorama was close to change with novel scientific endeavors. Namely the discovery of 5 new phyla within the archaea in less than 10 years

*Archaea diversity: new groups, new topologies.*

Already in 2007, extensive studies suggested sisterhood relation between thermoplasmatales and eukaryotes (Pissani *et al.,* 2007), supporting cytological and biochemical considerations previously noted by Margulis and Stolz regarding ultrastructural features (Margulis & Stolz 1984).

Before 2008, the differential distribution of eukaryotic homologues between crenarchaeal and euryarchaeal was considered evidence for chimeric scenarios where two archaea coevolved and then endosymbiosis occurred (see hydrogen hypothesis). In 2008, Yutin proposed that the crenarchaeal gene input to euryarchaea is significant, and that horizontal gen transfer can explain the chimeric distribution of protoeukaryotic genes within the Archaea (Yutin *et al.,* 2008).

The 2002 discovered Korarcheota phylum was genome-sequenced in 2008 (Huber *et al*., 2002; Lapidus *et al.,* 2008). The same year, marine mesophilic Thaumarcheota (Brochier-Armanet *et al.*, 2008) were identified. The phylogeny started to fill the gaps within the evolutionary relationship of crenarcheaota and eukaryotes. The new archaeal phyla were found in different habitats. Aigarchaeota were discovered in 2011 (Nunoura *et al.,* 2011). The new diversity allowed establishing sisterhood between Taumarcheaota, Aigarcheota, Crenarchetoanad and Korarcheota. This resulted in a new superphylum: the TACK archaea (Guy & Ettema, 2011).

*Resolving the tree of life in the postgenomic era*

Major advantage had been made since the serendipitous proposal of Woose (Woose *et al.*, 1990). Contemporary trees of life use 28 to 49 universal markers (Cox *et al.,* 2008; Foster *et al.,* 2009; Guy & Ettema, 2011). The new view is a two domain of life. Eukaryotes are part of a polyphyletic branch in archaeal phylogeny (Williams *et al.* 2013; Nesselquist & Gogarten 2013; Williams & Embley 2014; Guy & Ettema 2014). They are sisters of the TACK superphylum supporting previous suggestion about a chrenarchaeal origin (Williams & Embley 2014).

Massive gene gain was described at the basis of the TACK superphylum (suggesting that archaeal branches suffered reduction of genomes from a last common ancestor that had a complex genome (Wolf *et al* 2012). Reduction can be attributed large effective populations under strong selection pressures, as expected in extremophiles (Forterre, 1995). Complexity ca arose form explosive genome expansion in early eukaryotes, either close to LACA or TACK ancestor. (Wolf and Koonin, 2013).

In 2012 a phylogeny using 32 universal markers succeed to root the whole TACK with eukaryotes as they sister branch but failed to explain their fine relation within the TACK (Williams *et al.,* 2012; 2013). However, the picture of eukaryotic origin close to the TACK is consistent with the presence of informational homologs such as RNAse polymerase subunits RPB8 (Koonin *et al.,* 2007); RPC34 (Blombach *et al.,* 2009) and transcription factors (Daniels et al 2009). In contrast euryarchaeal homologs are mostly operational genes (Yutin *et al.,* 2012).

*Two-domain tree of life*

Super matrix approach in phylogenomics allowed avoiding long-branching attractions and politomies artifacts in 2010. Eukaryotes were placed next to poorly understood Korarchaeota (Gribaldo *et al.* 2010). Simonetta Gribaldo, on a conference on major transition of evolution held in Mexico City in 2015 argued about the importance to define topologies, not only in the TACK, but to improved super matrices at the root of bacteria and Archaea (Gribaldo, 2015). The consequences of rooting the tree of life are as we have seen fundamental for eukaryogensis (Poole & Penny, 2007). For example, the Archeozoa hypostasis could be plausible again if the phylogenetic tree of life can be single rooted. If only one domain exists, then gradualistic change from the bacteria could provide evidence for Cavalier-Smith's *Neomuran* revolution.

In 2015 new genomes assemblages form the Arctic Ocean were named Lockiarcheaota (Spang *et al*., 2015). They revealed to be the sister group of eukaryotes and solved the topology towards the other TACK members. Lockiarcheaota is only a metagenomics assemblage but has revealed to contain genes specific for membrane remodeling vesiculation and other typical eukaryotic features (Spang *et al*., 2015; Embley & Williams, 2015; Nasir *et al.,* 2015). The two-domain tree of life became relevant in the light of a rooted tree of the TACK superphylum: a scenario of gradualist speciation where eukaryotes are included.

**The dispersal prokaryotic Eukaryome**

Eugene Koonin has pointed out the concept of dispersal Eukaryome, regarding the extended distribution or eukaryotic homologs in Prokarya and the specific distribution in the different lineages of Archaea (Koonin, 2015). The classical perspective of searching cytological homology between eukaryotes and prokaryotes, which defined archeozoa and endosymbiotic theories, has moved to functional and comparative genomics of specific cell systems. Here I will summarize some of Koonin arguments about specific cell systems.

*Ubiquitin system*

Ubiquitin system was believed to be a eukaryotic feature evolved from prokaryotic enzyme synthesis. In 2011 ubiquitin putative homologs of archaea E1 and E2 were found to harbor operon structure in *Candidatus Caldiarchaeum subterraneum*, the only representative of Aigarchaeota (Nunoura *et al*, 2011). Further analysis found sequence similarity to eukaryotic E1 and E2 genes (Koonin, 2015). Also, the sulfur carrier homolog URM was found in Sulfolobales suggesting that the ubiquitin system is part of an early degradation pathway in the TACK (Marakova and Koonin, 2010).

*Cytoskeleton*

Cytoskeletal features were thought to be sinapomorphies of eukaryotes and a requirement of protoeukaryotic cells. Distant prokaryotic homologs of tubulin are found in septation proteins FtsZ and MreB. Comparative genomics revealed actin homologs in crenarchaeal Thermococcalees, Korarchaeota and Aigarchaeota. These proteins can coil in the same way as eukaryotic actins (Bernarder *et al.,* 2011). Recent tubulins homologs were found in Thaumarchaeota completing the picture of actin tubulin system as a probable feature of the TACK ancestor (Yutin and Koonin 2012).

*Cell division*

Mostly all prokaryotes share division mechanisms based on Z-ring formation by FtsZ. Some variants of cell division such as endospore formation and offspring viviparity (Enterobacteriales, Firmicutes and Myxococcales) have been described in endosymbionts but poorly characterized at the molecular levels. (Angert, 2015). Additionaly to FtsZ; a homolog on the ESCRT-III membrane remodeling complex has been found in crenarchaeal orders (Lindas *et al.,* 2008) and in some euryarcheota (Makarova *et al.,* 2010). The Thaumarchaeota *Nitrospumilis marins* uses ESCRT-III as primary cell division machinery (Pelve *et al.*, 2011). Comparative genomic analysis also provides evidence of a third system in thermoprotiales, which in the context of a dispersive distribution of ESCRT-III and FtsZ suggest that the three systems might have coexisted at the root of the TACK (Koonin, 2015). Furthermore, it has been appointed that cell cycle variation in prokaryotes led to genome expansion and polyploidy, as suggested for the giant bacteria *Epulopisicum fishelsoni* and the whole euryarchaeota clade (Katz and Oliverio, 2014). Thus, the picture of a progressive TACK archaea genome reduction and explosive genome expansion in eukaryotes and has gain much more discursive support.

*Endocytic system*

Studies on ESCRT in eukaryotic phylogeny have revealed that endocytic systems are diverse among eukaryotic groups. Endocytosis is mediated by novel proteins AP5 and TOM-1, which act on three different ESCRT systems (I; II; III) in different configurations depending on the eukaryotic major clade. This suggest that all variants of configuration of endocytic Pathway already existed in the LECA (Wideman *et al*., 2014). This could be consistent with a scenario where phagotrophy was an important trait that leads to adaptive radiation in early eukaryotic evolution.

*Nuclear envelope*

From the perspective of comparative biochemistry, it has been shown that the nuclear pore complex (NPC) and the endomembrane system could have coevolved. They are present in the form of a proto-coatomer in all major eukaryotic lineages suggesting that the LECA already had a NPC (DeGrasse *et al.*, 2009). Furthermore, multiple DNA binding domains have been described in major eukaryotic groups suggesting a diversity of membrane protein-nucleic acid interaction in the LECA (Devos *et al.,* 2006). Finally, the existence of multiple systems responsible for chromosome and plasmid segregation in bacteria suggests that chromosome segregation could have triggered nuclear membrane evolution and mitosis (Dawson &Wilson, 2015).

*Nucleolus*

Staub *et al*. searched for homologues of nucleolar protein domains based in the nucleolar proteome of S. *cerevisae*. Many protein domains were found in archaea (Staub *et al*., 2004). There are at least 25 nucleolar domains shared between arches and eukarya; 13 between prokarya and eukarya , exclusive of eukarya 25 and 29 present in the three domains . For example, enzyme domains and remodeling factors RNA bases are homologous to archaea. Among these, were discovered: Sm protein, present in the H subunit of RNA polymerase I (Hermann *et al.*, 1995) ; splicing factors ( U1 , U2 , U4 ) ; CBF transcription factor (Burley *et al.*, 1997) ; proteins ( S3A , S4 , L15 , L31 ) ; and ribosomal factors ( eIF - 5th , eIF-5ª_N , eIF6 , eRF1_1 , eRF1_2 , eRF1_3 ) used in the translation process ( Koonin, 1995 ). There are 29 exclusive domains of eukaryotes that have to do with fundamental functions of nucleolus ( Staub et al, 2004) . These include the HMG genes, ribosomal domains (Ribosomal_L6e, Ribosomal_L14e , Ribosomal_ L22e , Ribosomal_L27e ) ( Ghallagher *et al.* , 1994), the recognition domain of the SRP ( Birse *et al*. , 1997) , the PARP domains polymerases ( Smith, 2001 ) and chromatin remodeling complex CHROMO ( Koonin *et al*. , 2005) SSU complexome proteins were founded as homologues in archaea (Wen et al., 2013) and that the length of the intergenic spacer may affect nucleolar size and structure (Thiry et al., 2005).

*Interference RNA´s*

Regulation of gene expression through enhancers, transcription factors, and interference RNA might seem as an idiosyncratic invent of eukaryotes towards evolution of genome complexity. *Argonauts* families are found be originated in Euryarchaeota (Marakova *et al.,* 2009). *Dicers* don't have direct homologs, but some protein domains resembled homology between RNAse III and bacterial protein domains. The archaeal counterpart of Dicer helicase domain is a homolog in euryarchaeal helicase (Shabalina & Koonin, 2008) Thus interference RNA might be an idiosyncrasy of eukaryotes on the basis of protein domain adaptation to new genomic architecture.

According to Koonin the picture of a dispersal Eukaryome among the prokaryotes but with a clearly functional as phyletic pattern on the TACK superphylum, suggest that eukaryogensis took place in the root of the TACK. Genome expansion occurred at the root of the archaea and genome reduction at their branches. Thus, complexity might be related to complex genomes either contemporary to LACA or to the ancestor of the TACK (Wolf & Koonin, 2013; Koonin, 2015).

**The disciplinary debate**

The archeozoa hypothesis by Cavalier Smith, which originally suggested a bacterial origin of the archeozoa, can be dismissed by the fact that eukaryotic features as we know them today, were unsustainable at the complex forms of the eukaryotic ancestor, due to their unsustainability in the context of prokaryotic cells. A complex archaeon is a plausible model for gradual evolution of genome complexity and at some point, assimilation of endosymbiotic protobacteria as energy resources (Lane, 2011). Definitive topologies in phylogenetic hypothesis might reveal a better approach to the discrete relation of eukaryotes to the TACK however the so called phylogenomics impasse can lead to misunderstanding in teleological basis of the questions around eukaryogensis. On one hand the question about the abstract nature of the protoeukaryotic cell and on the other hand the question on an historic narrative of the gain of eukaryotic features from FECA to LECA. The last has been the source of controversy between the endosymbiotic theories and the idiosyncratic scenarios but also a sort of speculation discourse and tautology until recent years.

Margullis later theory on the origin of nucleus departs form the necessity of microbial evolution to include cytological data set in the molecular era. For her, Woose classification is partial, as the separation between archaea and bacteria from cytological point of view is as plausible as separation between archaea and eukaryotes. The last could have only emerged from several chimeric episodes due to their complexity level, but not form gradualistic process. Cytological characters have become more attention today in the context of better-known archaeal diversity. Also, symbiotic relations in microbial ecosystem became relevant for defining evolutionary traits in gradualistic as saltationst scenarios.

Although viral eukaryogensis is an attractive scenario it should regarded as the archeozoa hypothesis jut as an actualistic view to the problem of eukaryotic origin. Phylogenetic basis is required to prove homology between viral genomes and the other three domains. If a putative virome can resemble de dispersal nature of the Eukaryome, especially in those genes that are thought to be idiosyncratic, practical advantages can be made.

The ecosystem first approach is clearly much more theoretical than biological. Despite its tendency to autorganization theory, some aspects are important. First, that an ecosystem first approach can be implemented for the rise of major transitions of evolution in which the evolution of ecosystem is a fundamental part toward explaining evolution of species. And secondly that the debate between xenogenesis and idiosyncrasy is still open so far, the evolutionary traits at the root of the tree of life remain an open question. In my opinion so far, our methods cannot provide data to construct a sharp ecological causality, the complex system approach must remain as a metaphysical domain but not as a scientific theory.

In the paleobiolgical approach, just relative datation of LECA is possible yet. However, the paleobiological reconstruction of eukaryotes originated over chronical oxygenation in microbial mats ecosystem, where phagotrophy triggered cell size evolution and radiation is to be consider in exosystemic and metabolic perspectives.

A certainly important contribution has been de discovery and classification of archaeal diversity. From the modern phylogenomics methods a skeleton for historic narrative has been proposed. The TACK superphylum is now an open sea for scientific endeavor. Scientists are starting to view to the biology of those organism rather than to their genomes. We have moved from the reductionism of ribosomal phylogenies to the multiplicity of genomics and now to integrative biology. Therefore, a new disciplinary basis is needed.

**Evolutionary cell biology**

The *-omic's era* come with the challenge of relating sequence information with biological meaning of structure and function. The cell might be still a valid conceptual framework to address this interaction. As Lynch said, "*All aspects of biological diversification ultimately trace to evolutionary modifications at the cellular level" (Lynch, 2014).* The research program proposed by Lynch and collaborators in 2014 departs from the premise that a transdisciplinary interaction between cell biology and evolutionary science is possible to understand complex aspects of the major transitions in evolution. But it also challenges to encompass epistemic differences of both disciplines. Can we use the mechanistic nature of cell and molecular biology to explain the phenotypic diversity of cells? How can we integrate both disciplines if cell biology is reductionist in comparison to the integrative nature of evolutionary biology addressing diversity through historic narrative? One should recognise that evolutionary biology has become more reductionist and less speculative in terms of cell evolution due to the realm of well-resolved phylogenies. In this sense the cell biological approach does not conflict with the reductionistic nature of functional disciplines such as cell biology, but with the fact that cell studies are limited to few model organisms. In summary, if we want cell biology to be complementary to evolutionary inquiry, the approach must address diversity and retrieve a comparative approach. Therefore, Lynch *et al* proposal focuses on four questions (Lynch *et al.,* 2014):

- Why are cells the way they are and why aren´t they perfect?
- How do cellular innovations arise?
- Where do cellular innovations map onto the tree of life?
- How can evolutionary cell biology be effectively implemented?

The inquiry process in evolutionary cell biology departs from genomic and proteomic analysis, being based on functional constrains known from model organisms. Protein domain analysis has proven a powerful tool to address homology of functional entities. Mapping these characters on comprehensive phylogenies may reveal important insights on the biology of common ancestors. Then, functional studies can be conducted to prove specific putative functions in non-model organisms. Morphology plays two major roles: to establish better morphological characters according to new phylogenies, and to assess experimental studies on structure, topology, and localization of cell functions, related to phylogenies. The reconstruction of ancestral states of an organelle's architecture is an example (Lynch, 2014; Richardson *et al.,* 2015). As a result, evolutionary narratives will be able to rely on more statistical and functional evidence.

### *Eukaryogenesis from the perspective of evolutionary cell biology*

As we have seen efforts on eukaryotic origins have taken places form different disciplines and modes according to the historical curse of biology (Lynch & Field, 2014). We need to integrate these different approaches but not only in the discursive argumentation of biological essays, but in the design of experiments on functional genomics, cytological description and phylogenetic reconstruction. Thirty years earlier Doolittle argued towards a comprehensive understating of comparative cell physiology form evolutionary perspective (Doolittle, 1980; Hartwell *et al.,* 1999). Evolutionary cell biology should recognize the value of cytological and morphofunctional diversity of all eukaryotes, and therefore proposes research outside the canonical model organisms and turns to the tradition of compared cytology in an exercise that reconciles morphology, biological actualism and phylogenomics. Thus, new theories try to reconcile paradigmatic events as endosymbiosis, karyogenesis, the origin of cilia and endocytosis, in the light of the current evolutionary constrains of functional systems.

### *Cell topology across the FECA-LECA evolution*

A recent paper by Gabaldón & Pittis is a remarkable example that calculates stemlenghts for eukaryotic protein families along the FECA-LECA branch, gives striking evidence for the potential sequentiality in which organelles could arose: -*The very teleological question that has driven many of previous efforts*-. According to Gabaldon, LECA already harbored all eukaryotic features and mitochondria were the last to be incorporated. Endomembrane system, nucleus and non-proteinaceous features were acquired long before (Gabaldón & Pittis, 2016). Does this mean gradualist evolution is a more plausible scenario than chimeric origin? Comparative genome architecture and proteomics revealed eukaryotes as chimeric assemblages of archaeal and eubacterial genes and protein domains. Horizontal gene transfer is an explanation for chimerism but syntenial ancestry of genes from bacteria throughout archaea and eukarya must be taken in account. This is especially important to discern between functional genes acquired through HGT and linear evolving characters. One remarkable account is that eukaryotic proteins are related to cellular compartments and compared to their prokaryotic homologues; concluding that some early emerging organelles have chimeric prokaryotic ancestry. In this sense, the next approach could be considered to explore the cell biology of prokaryotic groups harboring eukaryotic protein families. Bacterial and Archaeal evolutionary cell biology is a promising field, expanding the comparative approach to prokaryotic cell architecture.

*Nuclear architecture and evolution*

Nuclear architecture has always been related to evolution by the question on the origin of the cell nucleus. Despite its complexity, previous attempts addressing evolutionary hypotheses stayed at the speculative level. The discovery that *Giardia lambia* (a flagellated protozoan and intestinal parasite with pediatric relevance) harbors a nucleolus by Jiménez-García et al in 2006 was for us the motivation to address the evolution of the nucleolus from a functional and evolutionary approach. The nucleolus is a multiproteic complex of close to 300 different proteins that are involved with processes as important as programmed cell death, metabolic regulation, cell differentiation, stress and aging (Boulon, *et al.,* 2010). All these processes converge to the core process occurring in the nucleolus, namely, the transcription and maturation and synthesis of ribosomal RNAs and components. The nucleolus should be considered a cell structure and a genomic location, therefore a nuclear domain that controls fundamental eukaryotic features relating them to the scale of ribosomal gene expression. Thus, the discovery of Giardia´s nucleoli implicated a major advance, namely that nucleoli represent a character shared among all eukaryotes (a sinapomorphy) and thus, LECA should also harbor a nucleolus.

*Nucleolus evolution*

The nucleolus has crenarchaeal and actinobacteria shared ancestry (Gabaldón & Pittis, 2016). I would like to conclude with the thought that non canonical organelles have been shown to be the early emerging features in eukaryogensis. Ribonucleoproteins and nuclear bodies are important to be consider as proteinaceous organelles or systems, harboring a large diversity of interactions and nanoscale morphologies (see Table 1).

The diversity of nucleolar morphology among eukaryotes and especially protists was already known since the 19[th] century based on strong cytological evidence (Montgomery, 1900, Heath, 1980). In fact, our knowledge on the mammalian and yeast nucleolus does not guarantee that the polymorphic structures of protists are homologous to these model nucleoli. Additionally, our definitions of nucleolus and nuclear bodies are undergoing reconsideration as the nanometric nature of protein-nucleic acid interaction and its relation to gene expression is being studied as a dynamic process (Mistelli, 2005). We face the challenge of understanding the evolution of processes of cell entities that are highly dynamic. We propose that the functional and structural knowledge on morphology, function, and development of Giardia´s nucleoli (Jiménez-García *et al*., 2006, Lara *et* a., 2016) and other early branching eukaryotes as *Entamoeba histolytica* (Vázquez-Echavarría *et al., 2009*), *Chaos chaos*, and *Pelomyxa palustris* (Islas-Morales et al, unpublished), increases our understanding of a 'primitive' nucleolus. Therefore, consistent evolutionary characters will necessary. Properties like, the behavior of nucleolus during mitosis, its distribution, and the presence of transcriptionally active polymerases have attracted our interest towards discrete elements of nuclear architecture that can justify characters at the evolutionary level. Thus, our current work focuses on applying an evolutionary cell biology approach with a focus on the emerging concepts of nuclear architecture.

## Conclusive remarks

One important lesson is to be learned from this review is that the question on the origin of eukaryotic cells demands trans-disciplinary work. Evolutionary cell biology is the product of a profound understanding of the epistemic aspects of evolutionary theory and its difference with functional approaches such as experimental genomics and cell biology. However evolutionary theory can become a practical unifying concept and theoretical background in the questions that evolutionary cell biology aims to address. For this understanding of evolution beyond adaptation must be communicated, but also experimental and theoretical methods for the early evolution of life must improve. Only then research projects can find a methodology of evolutionary narratives that can be supported by experimental evidence form cell biology. For eukaryogenesis immediate efforts must be driven to improve model organisms for lower eukaryotes and Archeaea. This will improve research on the origin of genome architecture, expression patterns and the compared cell biology of diverse cell systems.

## Acknowledgements

We want to thank professors: Valeria Souza from Instituto de Ecología at UNAM and Arturo Becerra from Facultad de Ciencias at UNAM, for their critical reading and comments. Parsifal Fidelio Islas-Morales developed part of this research as doctoral student from the ''Programa de Doctorado en Ciencias Biomédicas", Universidad Nacional Autónoma de Mexico (UNAM) and received a CONACyT fellowship 495217.